\documentclass[prl,twocolumn,superscriptaddress,floatfix]{revtex4}
\usepackage{CJK}
\usepackage{graphicx}
\usepackage{epstopdf}
\usepackage{color}
\usepackage{ulem}
\usepackage{gensymb}

\begin{document}
\begin{CJK*}{}{}

\title{Evidence of nodes in the order parameter of the superconducting doped topological insulator Nb$_x$Bi$_2$Se$_3$ via penetration depth measurements}
\author{M.~P.~Smylie}
\affiliation{Materials Science Division, Argonne National Laboratory, Argonne, IL 60439}
\affiliation{Department of Physics, University of Notre Dame, Notre Dame, IN 46556}
\author{H.~Claus}
\affiliation{Materials Science Division, Argonne National Laboratory, Argonne, IL 60439}
\author{U.~Welp}
\affiliation{Materials Science Division, Argonne National Laboratory, Argonne, IL 60439}
\author{W.-K.~Kwok}
\affiliation{Materials Science Division, Argonne National Laboratory, Argonne, IL 60439}
\author{Y.~Qiu}
\affiliation{Department of Physics, Missouri University of Science and Technology, Rolla, MO 65409}
\author{Y.~S.~Hor}
\affiliation{Department of Physics, Missouri University of Science and Technology, Rolla, MO 65409}
\author{A.~Snezhko}
\affiliation{Materials Science Division, Argonne National Laboratory, Argonne, IL 60439}

\date{\today}

\begin{abstract}
The low-temperature variation of the London penetration depth $\lambda (T)$ in the candidate topological superconductor Nb$_x$Bi$_2$Se$_3$ (x~=~0.25) is reported for several crystals.
The measurements were carried out by means of a tunnel-diode oscillator (TDO) technique in both field orientations ($H_{rf} \parallel$ $c$ and $H_{rf} \parallel$ $ab$ planes).
All samples exhibited power law behavior at low temperatures ($\Delta\lambda\sim T^2$) clearly indicating the presence of point nodes in the superconducting order parameter.
The results presented here are consistent with a nematic odd-parity spin-triplet $E_u$ pairing state in Nb$_x$Bi$_2$Se$_3$.
\end{abstract}

\maketitle
\end{CJK*}

Topological insulators, predicted and realized in the past several years \cite{Kane-Mele-PRL-Z2-topological-order-and-QSHE,Qi-RevModPhys-Review-of-TSC-theory,Hsieh-Hor-Cava-Nature-discovery-of-first-3D-TI,Hasan-Kane-RevModPhys-Review-of-TI}, are materials that display new quantum mechanical states that arise from the topology of their electronic structure.
In particular, gapless electronic surface / edge states arise whereas the bulk electronic structure is gapped.
The creation of topological states has been extended to topological superconductors \cite{Sasaki-Mizushima-PhysicaC-review-of-SC-doped-topological-materials,Mizushima-JPhysJap-Topological-SF-and-SC-review,Tanaka-JPhysSocJap-Review-Symmetries-in-TSC,Ando-Fu-RevCondMat-SOC} in which a superconducting bulk gap, complete or nodal \cite{Schnyder-Brydon-JPhysCondMat-nodal-topological-SC-review}, coexists with gapless surface states \cite{Qi-Zhang-PRB-TRI-SC-gapless-states,Sato-PRB-gapless-states-in-TSC,Hao-Lee-PRB-spectra-ABS-in-TSC,Hsieh-Fu-PRL-ABS-in-TSC-CuxBi2Se3,Fu-PRB-NematicCuxBi2Se3}.
Quasiparticle excitations of the surface states are considered Majorana fermions.  Additionally, as these surface states are topologically protected and thus robust against disorder, there is the potential for their application in fault-tolerant quantum computing \cite{Akhmerov-PRL-qbits-in-TIs-QuantumComputing, Wilczek-NatPhys-Majorana-QuantumComputing}.

The emergence of topological superconductivity depends sensitively on the material's symmetries: time reversal symmetry, spin rotation symmetry, inversion, and other crystal symmetries.
For instance, in a time-reversal symmetric and inversion symmetric system, the topological nature of the superconducting state is determined by the shape of the Fermi surface and the symmetry of the order parameter.
Specifically, odd-parity pairing, $\Delta$(-\textbf{k}) = -$\Delta$(\textbf{k}), and a Fermi surface containing an odd number of time reversal invariant momenta, \textbf{k} = -\textbf{k} + \textbf{G} with \textbf{G} a reciprocal lattice vector, will yield a topological superconductor.
In the case of weak spin-orbit coupling, odd-parity pairing corresponds to a spin-triplet pairing.
Thus, conventional s-wave superconductors are not topological and do not display Majorana surface states.
However, the coupling of the electron wavevector to its spin through strong spin-orbit coupling can induce unconventional pairing symmetries in time reversal symmetric systems~\cite{Fu-PRB-NematicCuxBi2Se3, Fu-Berg-PRL-CuxBi2Se3-as-candidate-TSC-theory,Ando-Fu-RevCondMat-SOC}.
In particular, a spontaneously broken spin-rotation symmetry is expected to arise at $T_c$ giving rise to a nematic state.

As the requirements for realizing topological superconductors are similar to those for creating topological insulators, namely strong spin-orbit coupling and electronic structures of specific symmetry, extensive work has been devoted to induce superconductivity in topological insulators by doping or applying pressure~ \cite{Sasaki-Mizushima-PhysicaC-review-of-SC-doped-topological-materials,Wang-Ando-ChemMat-TlxBi2Te3}.
Among these, Cu$_x$Bi$_2$Se$_3$ is the most studied~\cite{Hor-Cava-PRL-discovery-of-SC-in-CuxBi2Se3,Wray-Hor-Cava-NaturePhysics-Topological-order-observed-in-CuxBi2Se3, Schneeloch-Tranquada-CuxBi2Se3-high-volume-fraction,Wang-Jiang-NatSciRep-Electrochemical-CuxBi2Se3-Synthesis2016}.
An odd-parity, spin-triplet pairing state has been proposed \cite{Fu-PRB-NematicCuxBi2Se3,Hao-Yang-PRB-Spin-singlet-CuxBi2Se3-theory,Brydon-PRB-Odd-parity-SC-CuxBi2Se3, Nagai-PRB-impurities-and-CuxBi2Se3}, but not all data is consistent with this model.
Point-contact spectroscopy measurements of this material show a zero-bias conductance peak \cite{Sasaki-Ando-ZBCP-in-CuxBi2Se3,Kirzhner-ZBCP-in-CuxBi2Se3,Ando-Sasaki-JPhysConfSeries-More-evidence-of-ZBCP-in-CuxBi2Se3} indicating unconventional superconductivity, thermal transport measurements \cite{Kriener-Ando-PRL-CuxBi2Se3-specific-heat} are not fully consistent with the BCS model for a fully gapped system, and anomalies in the dc magnetization \cite{Das-Kadowaki-PRB-CuxBi2Se3-anomalous-magnetization-says-spin-triplet} indicate triplet pairing.
Furthermore, no Pauli limiting effect is observed in upper critical field measurements \cite{Bay-deVisser-PRL-CuxBi2Se3-pressure-study-upward-curvature-Hc2}, also supporting odd-parity spin-triplet pairing.
However, very low temperature ($\sim$15 mK) scanning-tunneling microscopy (STM) measurements \cite{Levy-PRL-15mK-STM-on-CuxBi2Se3} show a conventional, fully-gapped BCS-like s-wave structure.
Angular dependent NMR \cite{Matano-Kriener-Ando-NatPhys-SRS-breaking-in-CuxBi2Se3} and specific heat  measurements~\cite{Yonezawa} indicate the appearance of a two-fold in-plane anisotropy indicative of a nematic superconducting state \cite{Fu-PRB-NematicCuxBi2Se3,Fu-Berg-PRL-CuxBi2Se3-as-candidate-TSC-theory}.
Recently, superconductivity has been discovered in Nb-doped \cite{Qiu-Hor-NbxBi2Se3-arxiv-TRS-breaking-in-NbxBi2Se3} and Sr-doped \cite{Liu-Zhang-JACS-SC-discovery-in-SrxBi2Se3}  Bi$_2$Se$_3$. As in the case of the Cu-compound, the emergence below $T_c$ of a two-fold anisotropy in magnetotransport measurements on Sr$_x$Bi$_2$Se$_3$ \cite{Pan-deVisser-NatSciRep-SRS-breaking-in-SrxBi2Se3,GendaGu-HHWen-arxiv-Corbino-transport-in-SrxBi2Se3-2fold-symmetry} and in the in-plane magnetization of Nb$_x$Bi$_2$Se$_3$ \cite{Asaba-LuLi-arxiv-SRS-breaking-in-NbxBi2Se3} suggests the formation of a nematic state.

Here, we report on the first measurements of the low-temperature behavior of the superconducting penetration depth $\lambda$(T) carried out in both field orientations ($H_{rf} \parallel$ $c$ and $H_{rf} \parallel$ $ab$ planes) in single crystals of Nb$_x$Bi$_2$Se$_3$ with nominal composition of x = 0.25. The pairing symmetry is a key parameter in determining the topological state of the superconducting order which can be addressed with $\lambda(T)$ measurements. We find clear evidence for point-nodes in the superconducting gap in all crystals studied.
In conjunction with reports on a 2-fold in-plane magnetic anisotropy \cite{Asaba-LuLi-arxiv-SRS-breaking-in-NbxBi2Se3}, our findings are consistent with a nodal topological superconducting state of $E_u$ symmetry that preserves time reversal symmetry and breaks spin-rotation symmetry, as has been proposed in the ``nematic superconductor" model described by Fu \cite{Fu-PRB-NematicCuxBi2Se3}.
The critical field measurements yield a low superconducting anisotropy of $\gamma\approx$ 2, imposing constraints on the shape of the Fermi surface.

High-quality crystals of Nb$_x$Bi$_2$Se$_3$ (x = 0.25) have been grown by the same method used in Ref. \cite{Qiu-Hor-NbxBi2Se3-arxiv-TRS-breaking-in-NbxBi2Se3}.
The crystals show superconducting volume fractions approaching 100\%.
Nb$_x$Bi$_2$Se$_3$ crystallizes (Figure 1a, 1b) into the same tetradymite space group $R\bar{3}m$ as the parent material Bi$_2$Se$_3$ with a slightly extended c-axis due to incorporation of the Nb ion in the van der Waals gap between adjacent Bi$_2$Se$_3$ quintuple layers.
Crystals cleave easily between quintuple layers and naturally yield flat surfaces parallel to $ab$.

\begin{figure}
\includegraphics[width=1\columnwidth]{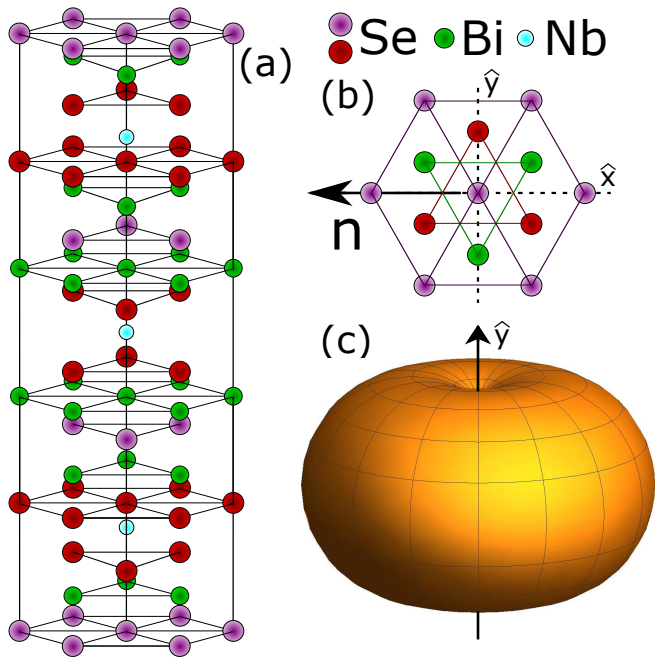}
\caption{
(a) Crystal structure of Nb$_x$Bi$_2$Se$_3$.
The Nb atoms are intercalated between quintuple layers of Bi and Se \cite{Qiu-Hor-NbxBi2Se3-arxiv-TRS-breaking-in-NbxBi2Se3}.
Some unit cells will only have two Nb ions.
(b) View along the $c$-axis, showing an axis of mirror symmetry $\hat{y}$; the nematic vector $\textbf{n}$ lies along a direction perpendicular to the mirror axis.
(c) Schematic of the superconducting gap structure in the $E_u$ state, where the nodal axis lies along $\hat{y}$.
}
\label{newfig1}
\end{figure}

Small crystals, approximately 850x700x150 $\mu$m in size, were first screened via SQUID magnetometry to identify superconductivity with the sharpest transitions.
Negligible variation in the onset of superconductivity around $T_c$~=~3.4 K was observed among crystals, and the transitions are sharp with the magnetization going flat at modest temperatures, indicating high-quality single-phase behavior.

Penetration depth measurements were carried out using a custom built \cite{BShen-Smylie-PRB-BaKFeAs-TDO,Smylie-PRB-BaFeAsP-TDO} 14.5 MHz tunnel-diode oscillator (TDO).
The samples were placed on a movable sapphire stage with temperature control from 0.4 to 30~K.
With this technique, the change in the resonator frequency $\Delta f(T)$ is proportional to the change of the London penetration depth $\Delta\lambda$,
\begin{equation}
\Delta f(T) = G\Delta\lambda(T) 
\end{equation}
where the geometrical factor \textit{G} depends on the sample shape and volume as well as the geometry of the resonator coil \cite{ProzorovGiannetta-SST-TDO-review}. The magnitude of the magnetic \textit{rf} field in the resonator coil used to sense changes in the penetration depth is  $\sim$20 mG, assuring that the sample remains fully in the Meissner state during measurements.

Figure 2 shows superconducting transitions of two Nb$_{0.25}$Bi$_2$Se$_3$ crystals as measured via the TDO technique with the \textit{rf} field applied along the $c$-direction of the crystals.
Curves were offset for clarity.
No secondary transitions indicating superconducting Nb ($T_c$~=~9.25 K) or NbSe$_2$ ($T_c$~$\simeq$~7.2 K) regions were observed at higher temperatures either via SQUID magnetometry or via TDO measurements.

\begin{figure} [b]
\includegraphics[width=1\columnwidth]{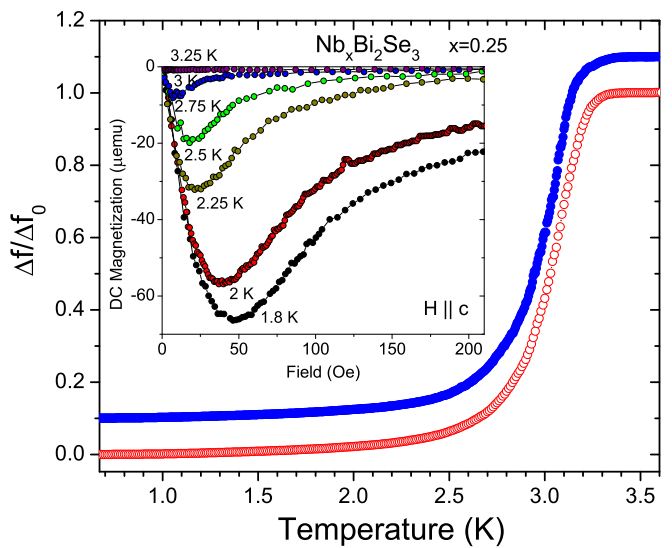}%
\caption{
(Color online) Temperature dependence of the normalized frequency shift in two select crystals of Nb$_x$Bi$_2$Se$_3$ with nominal doping x~=~0.25.
One of the curves has been offset for clarity.
The inset shows $dc$-SQUID magnetization curves versus field at temperatures from 1.8 K through $T_c$.
}
\label{fig1}
\end{figure}

To determine the superconducting anisotropy, the TDO frequency shift has been measured on a rectangular, plate-like single crystal with $dc$ fields ($H_{DC}$) applied both along the \textit{c}-axis as well as parallel to the \textit{ab}-plane.
The results are presented in Figure 3.
The magnetic phase diagram shown in Figure 3c was generated by determining the superconducting onset as the point where the TDO response had shifted 1 Hz below the normal state behavior.
Field-dependent measurements on another crystal with the \textit{c}-axis perpendicular to $H_{DC}$ were also performed, and the results of the two sets of measurements align.
Extrapolations to zero temperature give the values $B_{c2\perp}$(0 K) $\approx$ 0.9 T and $B_{c2\parallel}$(0 K) $\approx$ 1.8 T.
In-plane and out-of-plane coherence lengths of $\xi_{ab} \approx$ 19 nm and $\xi_c \approx$ 9.5 nm were estimated using the single-band Ginzburg-Landau relations $B_{c2\parallel} = \Phi_0 / 2\pi\xi_{ab}^2$ and $B_{c2\perp} = \Phi_0 / 2\pi\xi_c\xi_{ab}$.
Thus, Nb$_x$Bi$_2$Se$_3$ is characterized by a rather low superconducting anisotropy of $\gamma \approx~$2, similar to reported values of 1.5 and 1.8 for the Sr-doped and Cu-doped materials, respectively~\cite{Shruti-PRB-Gamma-in-SrxBi2Se3,Kriener-Ando-PRL-CuxBi2Se3-specific-heat}.
This low $B_{c2}$ anisotropy is indicative of an essentially 3D-electronic structure, consistent with recent quantum oscillation measurements on Nb$_x$Bi$_2$Se$_3$ \cite{x15}, that reveal a large ellipsoidal Fermi surface sheet and, in addition, smaller pockets, possibly derived from Nb $d$-orbitals.
At low fields, there is an anomalous upward curvature of $B_{c2\parallel}(T)$ that is not reported \cite{Bay-deVisser-PRL-CuxBi2Se3-pressure-study-upward-curvature-Hc2} in the homologue Cu$_x$Bi$_2$Se$_3$, but is observed \cite{Pan-deVisser-NatSciRep-SRS-breaking-in-SrxBi2Se3} in Sr-doped Bi$_2$Se$_3$. 

\begin{figure}
\includegraphics[width=1\columnwidth]{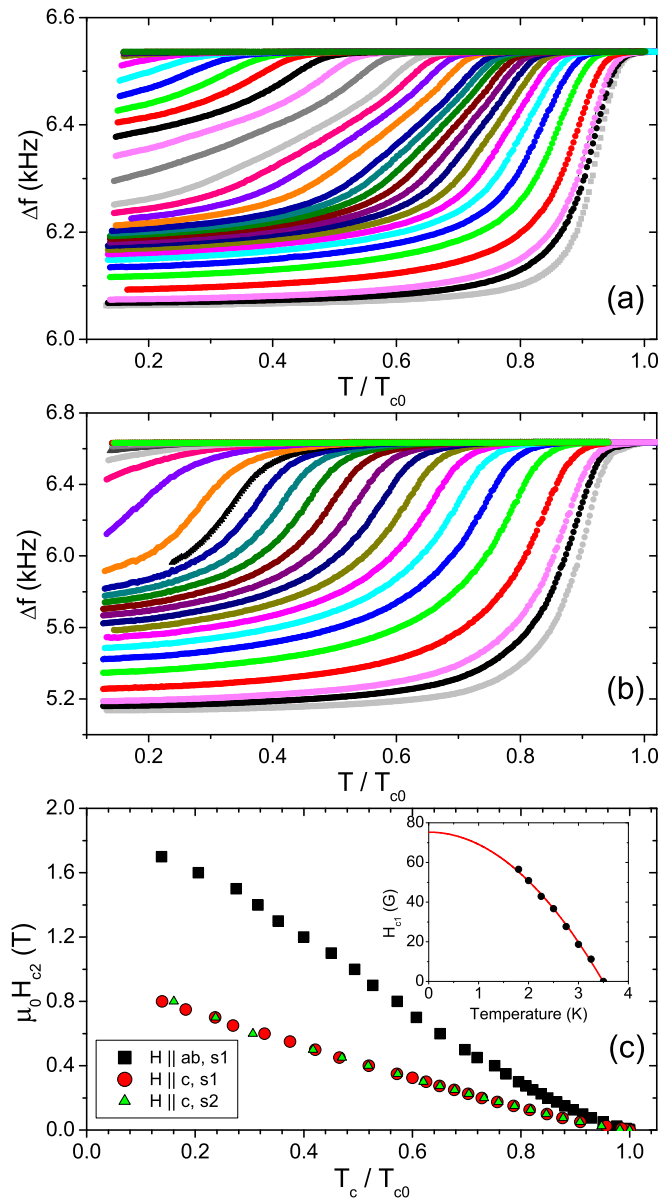}%
\caption{
(Color online) Field dependence of the TDO frequency shift $\Delta f(T)$ in a single crystal of Nb$_x$Bi$_2$Se$_3$.
(a) shows the suppression of $T_c$ with the c-axis perpendicular to $H_{rf}$.
(b) illustrate the suppression of $T_c$ with the c-axis parallel to $H_{rf}$.
(c) temperature dependence of the $H_{c2}$ field from the TDO data for two orientations.
Measurements on an additional crystal with $c$-axis parallel to $H_{rf}$ yield an identical phase curve (triangles).
The inset shows $H_{c1}$ values as extracted from $dc$ SQUID magnetization measurements.
}
\label{fig3}
\end{figure}

The inset of Fig. 2 shows the field dependence of the $dc$ magnetization measured at various temperatures in fields parallel to \textit{c}.
At low fields, the magnetization is linear in field as expected for the Meissner state.
With increasing field, deviations from linearity arise at a field $H_p$, signaling the penetration of vortices.
Since the sample is a plate with rectangular cross section, effects due to the geometrical barrier arise for which the relation of $H_p$ and $H_{c1}$ is given as $H_p /H_{c1}= $ tanh$ (\sqrt{\alpha t/w})$, where \textit{t} and \textit{w} are the thickness and width of the sample, and $\alpha$~=~0.67 for a disk \cite{Brandt3}.
At 2 K we find $H_p \approx$~18 Oe corresponding to $H_{c1} \approx$~51 Oe.
The inset in Fig. 3c summarizes the temperature dependence of $H_{c1}$ and a fit according to the Ginzburg-Landau prescription  $H_{c1}(T)=H_{c1}(0)[1-(T/T_c)^2]$ yielding $H_{c1}(0) \approx$~75 G.
We note that this fit allows for a rough estimate of $\lambda$(0) which is otherwise difficult to obtain, but does not account for the precise temperature variation of $\lambda$ which depends on the order parameter symmetry and is described in detail below.
Using the Ginzburg-Landau relation $H_{c1}=\Phi_0 / (4\pi\lambda_{ab}^2)$(ln$~[\lambda_{ab}/\xi_{ab}] + 0.5)$ we estimate $\lambda$(0) $\approx$ 237 nm and, with $\xi_{ab} \approx$ 19 nm, a Ginzburg-Landau parameter of $\kappa \approx$ 12.5, identifying Nb$_x$Bi$_2$Se$_3$ as extreme type-II.
We note that as $\lambda$(0) is rather large, the TDO measurements probe the gap symmetry in the bulk, not of a potential surface state.

The low-temperature variation of the penetration depth is determined by the distribution of the thermally excited quasiparticles on the Fermi surface, and by electron scattering.
A complete superconducting gap yields at sufficiently low temperatures an exponential variation of $\lambda(T)$, which in conventional BCS theory for an isotropic $s$-wave superconductor is given as
\begin{equation}
	\frac{\Delta\lambda(T)}{\lambda(0)} \approx \sqrt{\frac{\pi\Delta_0}{2T}}~\text{exp} \left(-\frac{\Delta_0}{T}\right)
\end{equation}
where $\Delta_0$ is the zero temperature value of the energy gap.
In contrast, gap nodes induce enhanced thermal excitation of quasiparticles, typically resulting in a power law variation of $\lambda$: $\Delta\lambda \sim T^n$ \cite{ProzorovGiannetta-SST-TDO-review,Gross-Hirschfeld-ZfPhysB-T2-lambda-in-UB13-d-vector-pinning}.

\begin{figure}
\includegraphics[width=1\columnwidth]{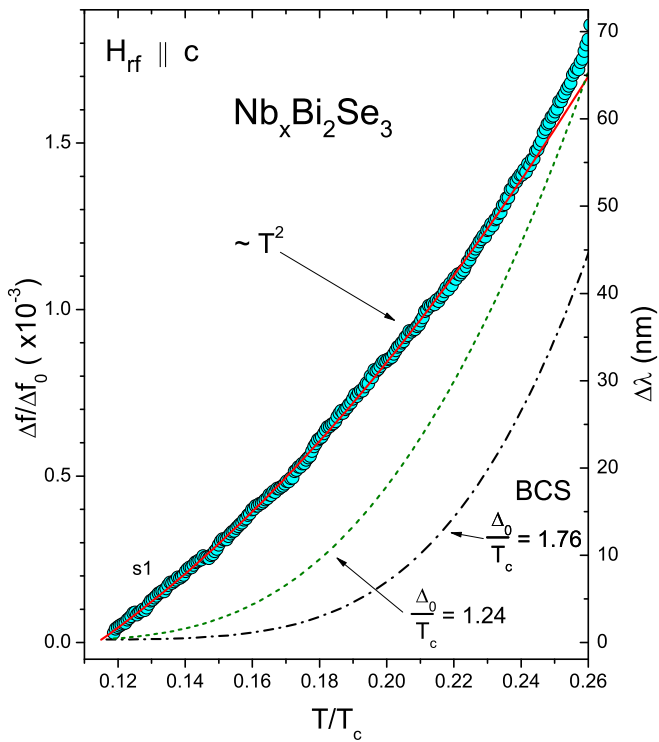}%
\caption{
Normalized low temperature frequency shift $\Delta f(T)$ in a single crystal of Nb$_x$Bi$_2$Se$_3$ plotted versus reduced temperature $T/T_c$.
Over a wide temperature range, the response is best described with a $T^2$ fit (solid line).
A fully-gapped BCS-like fit (black, dash-dotted curve) and a BCS-like fit with a free gap parameter (green, dashed curve) are plotted for comparison.
}
\label{figLowTTDO}
\end{figure}

Figure \ref{figLowTTDO} shows the low-temperature behavior of the relative TDO frequency shift  for a single crystal of Nb$_x$Bi$_2$Se$_3$ with $H_{rf} \parallel$ c along with several fits.
The standard BCS form with the weak-coupling gap value $\Delta_0$/$T_c$=1.76, is shown as a black dash-dot line (the fit was carried out to $T_c$/3). It clearly provides an inadequate description of the observed behavior.
A BCS-fit with a gap ratio $\Delta_0/T_c$ as a free parameter (dashed green line) within the same fitting range does not yield an adequate description of the data either.
This implies that the data do not represent a fully gapped superconductor with a gap that is significantly smaller than our measurement temperature.
The best fit to our data is achieved with a power law behavior with an exponent $n$=2.

\begin{figure}[t]
\includegraphics[width=1\columnwidth]{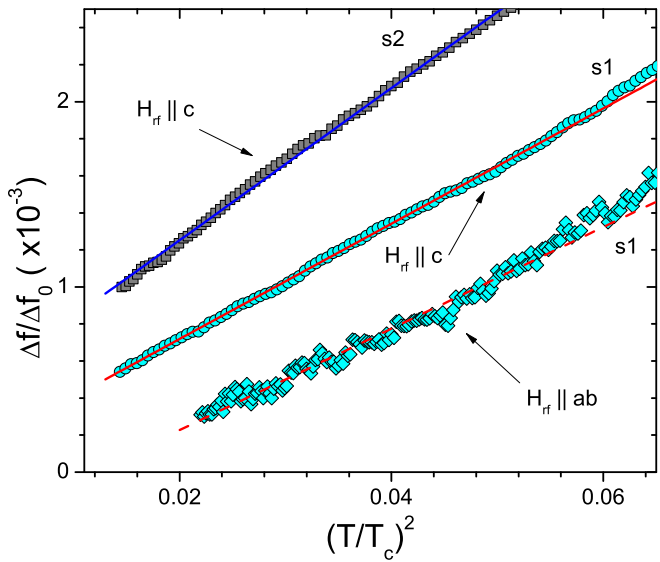}%
\caption{
Normalized low temperature frequency shift $\Delta$f(T) in a single crystal (s1, blue) of Nb$_x$Bi$_2$Se$_3$ with $H_{rf}\parallel$ c and $H_{rf}\parallel$ ab, plotted versus reduced temperature T/$T_c$ squared.
The dashed line is a quadratic fit of the data.
All data sets can be well fit with a straight line, indicating T$^2$ behavior in both orientations. An additional sample (s2, gray) also shows a $T^2$ dependence.
}
\label{figT2TDO}
\end{figure}

Low-temperature TDO frequency shift data for $H_{rf}~\parallel$~\textit{ab} measured on one of the samples is  shown in Fig.~\ref{figT2TDO} as a function of $(T/T_c)^2$ together with data for $H_{rf}~\parallel$~\textit{c}.
The total frequency shift in the parallel orientation is very small and the data is correspondingly noisier as in this orientation, the effective filling factor of the TDO sensor coil is much smaller.
All data reveal a quadratic low-temperature variation of $\lambda$.
This quadratic temperature dependence is consistent with linearly vanishing point nodes in the gap such as in the axial \textit{p}-wave spin-triplet state as has been reported previously for UBe$_{13}$ \cite{Einzel-Hirschfeld-T2-lambda-in-UBe13-says-pwave,Gross-Hirschfeld-ZfPhysB-T2-lambda-in-UB13-d-vector-pinning} or Sr$_2$RuO$_4$ \cite{Bonalde-vanHarlingen-PhysicaC-lambda-Sr2RuO4}, or in unconventional gap structures of topological superconductors \cite{Venderbos-Fu-arxiv-either-TRB-or-RSB-not-both} (see below).
In contrast, linearly vanishing line nodes, as encountered in the polar \textit{p}-wave state or in \textit{d}-wave superconductors, would yield a linear temperature dependence of $\lambda$.
Since the gap structure is anisotropic, different temperature dependences of $\sim T^2$ and $\sim T^4$ are expected, depending on whether the applied field is parallel or perpendicular to the symmetry axis of the gap \cite{Gross-Hirschfeld-ZfPhysB-T2-lambda-in-UB13-d-vector-pinning}.
Furthermore, in an anisotropic material the in-plane penetration depth $\lambda_{ab}$ and the interlayer penetration depth $\lambda_c$ are different and could in principle have different $T$-dependencies.
Formally, the observation of a $T^2$ dependence would in our geometry ($H_{rf} \mid\mid c$) correspond to an in-plane symmetry axis of the order parameter.
However, inhomogenous order parameter textures and/or inhomogenous field distributions brought about for example by sample edges lead to a mixture of components such that low-T measurements will be dominated by the $T^2$ term \cite{Gross-Hirschfeld-ZfPhysB-T2-lambda-in-UB13-d-vector-pinning}.
This could account for our observation that measurements for $H_{rf} \mid\mid c$ and $H_{rf} \mid\mid ab$ yield the same $T^2$ variation.
In addition, for $H_{rf}~\mid\mid c$, the TDO signal arises from in-plane currents that probe $\lambda_{ab}$.
When $H_{rf} \mid\mid ab$, the currents have in-plane and out-of-plane components implying that a mixture of both $\lambda_{ab}$ and $\lambda_c$ is probed.
However, the relative contribution of inter-plane penetration depth is roughly proportional to $(\lambda_c  t)/(\lambda_{ab}  w)$, where $t$ is the thickness of the sample and $w$ is its width~\cite{ProzorovGiannetta-SST-TDO-review}.
For our samples $t/w \approx 0.2$ and $\lambda_c/\lambda_{ab} \approx 2$ so the signal for both field orientations will be governed by $\lambda_{ab}$.

Our observation of point nodes in the gap of Nb$_x$Bi$_2$Se$_3$ is consistent with recent findings based on specific heat measurements \cite{Asaba-LuLi-arxiv-SRS-breaking-in-NbxBi2Se3}, which rule out line nodes but were not able to distinguish between point nodes and a complete gap.

The axial $p$-wave state contains a chirality and thus breaks time reversal symmetry.
In a recent extensive theoretical study \cite{Venderbos-Fu-arxiv-either-TRB-or-RSB-not-both} of odd-parity superconductors in trigonal and hexagonal crystal systems it was found that strong spin-orbit coupling produces either a time-reversal breaking, rotational symmetry preserving chiral phase (analogous to the axial $p$-wave state) or a time-reversal symmetric but spin-rotational symmetry breaking nematic state, both of which can be nodal.
A classification of possible time reversal symmetric pairing states in the $D_{3d}$ point group, applicable to Nb$_x$Bi$_2$Se$_3$, reveals that the two-dimensional odd-parity $E_u$ state spontaneously breaks the 3-fold in-plane rotational symmetry leading to a nematic state. This nematic state is characterized by the nematic director marking an in-plane direction of zero total spin.
For a director pointing in the $k_x$ direction an anisotropic fully gapped state with gap minima along $k_y$ arises, whereas for a director along $k_x$, a nodal state with point nodes along $k_y$ appears.
These nodes are symmetry protected owing to the mirror symmetry around the y-axis (see Fig. 1b).
Thus, our finding of point nodes in the gap in conjunction with the observation of a two-fold in-plane magnetic anisotropy \cite{Asaba-LuLi-arxiv-SRS-breaking-in-NbxBi2Se3} identifies the nodal $E_u$ state as the pairing state of Nb$_x$Bi$_2$Se$_3$.

Strong electron scattering may have profound effects on the temperature dependence of the penetration depth. For instance, impurity scattering can alter the linear T-dependence characteristic of line nodes to a quadratic variation as has been discussed for a variety of materials \cite{Hirschfeld-Goldenfeld-PRB-T*-scattering,x12,snezhko2004,x14,carrington}.
This point deserves further investigation, but we do not believe that our samples are in the dirty limit.
The observation of quantum oscillations in the Cu-homologue \cite{Lawson-Hor-LuLi-PRL-CuxBi2Se3-QuantumOscillations,Lawson-Li-PRB-2D-Fermi-surface-in-CuxBi2Se3} and more recently in the Nb-compound \cite{x15} suggests that these materials have in fact fairly high purity.
$T_c$ of superconductors with sign-changing order parameters is rapidly suppressed by impurity scattering \cite{y1, y2}.
For instance, particle irradiation experiments on nodal BaFe$_2$(As$_{1-x}$P$_x$)$_2$ \cite{Smylie-PRB-BaFeAsP-TDO,x14} showed a $T_c$-reduction by 20-25\% by the time the quadratic $T$-dependence of lambda is established due to irradiation induced defects.
In contrast, there is little variation in $T_c$ values among Nb$_x$Bi$_2$Se$_3$ crystals, implying that large sample-to-sample variations in impurity content are not significant.

In conclusion, we present the first measurements of the low-temperature penetration depth of high-quality single crystals of the candidate topological superconductor Nb$_x$Bi$_2$Se$_3$ (x = 0.25).
On multiple samples and for both directions of applied field ($H_{rf} \parallel$ $c$ and $H_{rf} \parallel$ $ab$) we find a quadratic temperature dependence $\Delta \lambda \sim T^2$ indicative of point nodes in the superconducting gap.
Our results are consistent with a nematic $E_u$ pairing state in which symmetry protected point nodes appear along the $k_y$ direction on the Fermi surface.
Exploring the origin of the differences between this material and the fully gapped homologues Cu$_x$Bi$_2$Se$_3$ \cite{Levy-PRL-15mK-STM-on-CuxBi2Se3} and Sr$_x$Bi$_2$Se$_3$ \cite{Du-HHWen-arxiv-Full-gap-in-SrxBi2Se3-via-STM} will be important in understanding the rich physics that arises due to strong spin-orbit coupling.

\textit{Acknowledgments --}
Tunnel diode oscillator and magnetization measurements were supported by the U.S. Department of Energy, Office of Science, Basic Energy Sciences, Materials Sciences and Engineering Division, Contract No. DE-AC02-06CH11357.
MPS thanks ND Energy for supporting his research and professional development through the ND Energy Postdoctoral Fellowship Program.
YSH acknowledges support from National Science Foundation grant number DMR-1255607.

\bibliographystyle{apsrev4}

\end{document}